\documentclass[12pt]{article}
\usepackage{pst-plot,epsf}
\newcommand{\beq}{\begin{equation}}
\newcommand{\eeq}{\end{equation}}
\newcommand{\bea}{\begin{eqnarray}}
\newcommand{\eea}{\end{eqnarray}}
\newcommand{\nobody}{\rule{0ex}{1ex}}

\newcommand{\epm}{e^+e^-}
\newcommand{\gv}{\mbox{GeV}}

\newcommand{\ra}{\rightarrow}
\newcommand{\ga}{\gamma}

\begin{document}
\thispagestyle{empty}
\begin{flushright}
DESY 00-164\\
December 2000\\
\vspace*{1.5cm}
\end{flushright}
\begin{center}
{\LARGE\bf Fermion mass effects in $e^+ e^- \ra 4f$ 
           and $e^+ e^- \ra 4f\gamma$ with cuts\footnote{Work supported 
           in part by the Polish State Committee for Scientific Research 
           (KBN) under contract No. 2~P03B~004~18.}}\\
\vspace*{2cm}
Fred Jegerlehner$^{\rm a}$ and
Karol Ko\l odziej$^{\rm b}$\vspace{0.5cm}\\
$\nobody^{\rm a}${\small\it
Deutsches Elektronen-Synchrotron DESY, Platanenallee 6, D-15768 Zeuthen,
Germany}\\
$\nobody^{\rm b}${\small\it
Institute of Physics, University of Silesia, ul. Uniwersytecka 4,
PL-40007 Katowice, Poland}
\vspace*{3.5cm}\\
{\bf Abstract}\\
\end{center}
The fermion mass effects in $e^+ e^- \ra 4f$ and in the corresponding
bremsstrahlung reactions in the presence of realistic cuts are
studied.  It is shown that, for some four--fermion final states, the
mass effects become sizeable to the extent that they may affect
the accuracy of theoretical predictions which is required to be better
than 1\%.
\vfill

PACS: 12.15.-y, 13.40.Ks   
\\
Keywords: Electroweak interactions, Electromagnetic corrections to weak-interaction processes   
\\

\newpage
\section{Introduction}

An increasing precision of the measurements of $W$-boson-pair
production at LEP2 and the prospect of improving it at high energy linear
colliders requires a corresponding high precision in the Standard
Model (SM) theoretical predictions. In order to match the requirements
of the final LEP2 data analysis, it is necessary to provide
theoretical predictions at a precision better than 1\% \cite{GP}. The
expected experimental precision of future $\epm$ colliders will be
probably about 0.1\%. Such high precision of the theoretical
predictions can only be achieved by including the complete set of 
electroweak radiative corrections at the one-loop level as well as the
leading electroweak logarithms at higher-loops of the Standard Model.

A calculation of the complete set of the SM one-loop virtual corrections
to $\epm \ra 4f$, which are reactions of actual interest, is a very tedious 
task, and despite the fact that some progress in calculating the 
corrections to $e^+ e^- \ra u \bar{d} \mu^- \bar{\nu}_{\mu}$ was
reported in Ref.~\cite{AV}, at present there is no final result available 
for any of the possible four--fermion final states. 
Fortunately, a theoretical precision which is satisfactory for most
applications in the analysis of the LEP2 data, can be achieved within 
the double-pole approximation (DPA) \cite{DPA}. Recently, an interesting
complete analysis of the virtual and real photonic corrections in the 
DPA has been reported in Ref. \cite{DDRW2}. In Ref. \cite{DDRW2}, the DPA 
has been applied actually only to the non-leading virtual $\cal{O}(\alpha)$ 
corrections while real photonic corrections have been obtained with the 
full matrix elements of $\epm \ra 4f\gamma$ in the massless fermion limit.
A great advantage of the double-pole approximation
is that its basic ingredients such as the production of the on-shell
W-pairs \cite{rc} and W-boson decay \cite{decay} have already been calculated
at one-loop in the SM. Also the so called non-factorizable virtual
corrections are known to have a simple structure and can be found in
the literature \cite{nf}.

Real photon corrections and in particular the hard bremsstrahlung are  
inherent ingredients of the $\cal{O}(\alpha)$ 
radiative corrections to the four--fermion processes and precise treatment 
of them is crucial for the ultimate accuracy of theoretical predictions,
which for a proper analysis of the final LEP2 data, should possibly match 
the level of 0.5\% \cite{GP}. At the moment, there exist several packages 
which allow one to calculate cross sections of $\epm \ra 4f\gamma$ for any 
possible final state. They have been compared in Ref. \cite{GP}.
Three of the codes based on full matrix elements: {\tt WRAP, RacoonWW 
and PHEGAS/HELAC} have been subject to tuned comparisons in the approximation 
of massless fermions in the presence of cuts and they show a very good 
agreement for the final states considered in Ref. \cite{GP}.
A complete list of results for total cross sections of all representative 
processes $\epm \ra 4f\gamma$ can be found in Ref. \cite{DDRW1}.


The fermion mass effects for $\epm \ra 4f$ and $\epm \ra 4f\gamma$ 
for different $CC10$ final states have been studied in Ref. \cite{JK}
for the total cross sections without cuts, except for the photon energy
cut. It has been shown that the cross sections of $\epm \ra 4f\gamma$ 
differ by up to a few per cent for final states including particles 
with different masses. The natural explanation for it is that the
collinear divergences are regularized by different fermion masses in
each case and therefore a small change in the fermion mass results
in sizeable effects in the total cross sections. One would not expect 
any numerically sizeable effects of the 
fermion masses in the presence of angular and invariant mass cuts, as 
the angles between particles are relatively big then.
However, as it will be demonstrated in the following, the massless 
fermion limit which is usually used in the non-collinear 
phase space region, may be a source of inaccuracy which may substantially 
affect the desired accuracy level of 0.5\%.

\section{The calculation}
In the present section we sketch the basics of the calculation.
We refer to Refs.~\cite{KZ} and \cite{JK} for more details.

The matrix elements of the reactions considered in the present paper
are calculated with the helicity amplitude method. Parts of the Feynman 
graphs which contain a single uncontracted
 Lorentz index are defined as generalized polarization 
vectors and used as building blocks of the amplitudes. Particular care is 
taken of the photon radiation off the external fermion lines. The 
corresponding fermion propagators are expanded in the light fermion
mass analytically in order to avoid numerical cancellations.
Fermion masses are kept non zero both in the kinematics and in
the matrix elements. Keeping the non zero fermion masses allows for
the proper treatment of the collinear photons. Therefore cross sections 
can be calculated independently of angular cuts and the background from 
undetected hard photons can be estimated.
Moreover, a photon exchange in the $t$-channel can be handled properly
and the Higgs boson effects can be incorporated consistently.

The photon propagator is taken in the Feynman gauge and the propagators 
of the massive gauge bosons $W^{\pm}$ and $Z^0$, are defined in the unitary 
gauge. The constant widths $\Gamma_W$ and  $\Gamma_Z$ are introduced 
through the complex mass parameters
\bea
\label{cmass}
M_V^2=m_V^2-im_V\Gamma_V 
\eea
in the propagators. However, the electroweak mixing parameter is kept real,
although there is no obstacle to having it complex. This simple prescription 
preserves the electromagnetic gauge invariance with
the non-zero fermion masses, even when the widths $\Gamma_W$ and 
$\Gamma_Z$ are treated as two independent parameters, which has been checked 
numerically, and for some final states, also analytically.

The constant width prescription violates the $SU(2)$ gauge-symmetry.
However, the corresponding numerical effects caused by spoiling the gauge 
cancellations are in the presence of cuts,  practically irrelevant up to 
the relatively high c.m.s. energy of 10 TeV. 
This observation relies on the comparison with the results
of Ref.~\cite{DDRW1}. Our results for $\epm \ra u\bar{d}e^-\bar{\nu}_e$,
$\epm \ra u\bar{d}\mu^-\bar{\nu}_{\mu}$ and the corresponding 
bremsstrahlung reactions, which were calculated in the linear
gauge, agree within statistical errors with those of Ref.~\cite{DDRW1}
which were obtained in a nonlinear gauge in the so called complex-mass
scheme that preserves the Ward identities. 

The phase space integration is performed numerically.
The 7 (10) dimensional phase space element of $e^+e^- \ra 4f$
($e^+e^- \ra 4f\gamma$) is parametrized in several different ways, which 
are combined in a single multichannel Monte Carlo (MC) integration routine.
In the soft photon limit, the photon phase space is integrated analytically.

The most relevant peaks of the matrix elements, e.g.,
the collinear peaking related to the initial and final state radiation,
the $\sim 1/t$ pole caused by the $t$-channel photon-exchange,
the $\sim 1/k$ peaking of the bremsstrahlung photon spectrum,
the Breit--Wigner shape of the $W^{\pm}$ and $Z^0$ resonances,
the $\sim 1/s$ behavior of a light fermion pair production,
and the $\sim 1/t$ pole due to the the neutrino exchange 
are mapped away before applying the MC integration routine {\tt VEGAS} 
\cite{vegas}.

\section{Numerical results}
In this section, we will present numerical results for several different 
four--fermion reactions $e^+e^- \ra 4f$ and the corresponding bremsstrahlung 
reactions.

We define the set of physical parameters, as in Ref.~\cite{GP},
by the gauge boson masses and widths:
\beq
\label{masses1}
m_W=80.35\; {\rm GeV}, \quad \Gamma_W=2.08699\; {\rm GeV}, \qquad
m_Z=91.1867\; {\rm GeV}, \quad \Gamma_Z=2.49471\; {\rm GeV}\; ,
\eeq
by the couplings which are defined in terms of the electroweak mixing 
parameter $\sin^2\theta_W=1-m_W^2/m_Z^2$ and the fine structure constant
at two different scales, $\alpha_W$ and $\alpha$, the latter being used for 
parametrization of couplings of the external photon,
\beq
\label{coups1}
\alpha_W=\sqrt{2} G_{\mu} m_W^2 \sin^2\theta_W/\pi,\qquad
        \alpha=1/137.0359895,
\eeq
with $G_{\mu}=1.16637 10^{-5}{\rm GeV}^{-2}$.

For the sake of comparison with Ref.~\cite{DDRW1} we introduce the second 
set of parameters, originally defined in Ref. \cite{Boudjema}, which are 
given again by the gauge boson masses and widths
\beq
\label{masses2}
m_W=80.23\; {\rm GeV}, \quad  \Gamma_W=2.0337\; {\rm GeV}, \qquad
m_Z=91.1888\; {\rm GeV}, \quad \Gamma_Z=2.4974\; {\rm GeV},
\eeq
the single value of the fine structure constant and the electroweak 
mixing parameter 
\beq
\label{coups2}
\alpha_W=1/128.07, \qquad \sin^2\theta_W=\pi/(\sqrt{2}\alpha_W G_{\mu} m_W^2),
\eeq
with $G_{\mu}=1.16639^{-5} {\rm GeV}^{-2}$.

The charged lepton masses are given by
\beq
\label{mlepton}
m_e=0.51099906\; {\rm MeV}, \quad m_{\mu}=105.658389\; {\rm MeV},\quad
m_{\tau}=1777.05\; {\rm MeV}
\eeq
and for the quark masses we take
\beq
\label{mquark}
m_u=5\; {\rm MeV}, \quad m_d=10\; {\rm MeV}, \quad m_s=150\; {\rm MeV}, \quad
m_c=1.5\; {\rm GeV}, \quad m_b=5\; {\rm GeV}.
\eeq

We define a set of cuts, identical to those of Ref.~\cite{DDRW1} 
\beq
\begin{array}[b]{rlrlrl}
\theta (l,\mathrm{beam})> & 10^\circ, & \qquad
\theta( l, l^\prime)> & 5^\circ, & \qquad 
\theta( l, q)> & 5^\circ, \\
\theta (\ga,\mathrm{beam})> & 1^\circ, &
\theta( \ga, l)> & 5^\circ, & 
\theta( \ga, q)> & 5^\circ, \\
E_\ga> & 0.1\;\gv, & E_l> & 1\;\gv, & E_q> & 3\;\gv, \\
m(q,q')> & 5\;\gv\,,
\end{array}
\label{cancuts}
\eeq
where $l$, $q$, $\ga$, and ``beam'' denote
charged leptons, quarks, photons, and the beam (electrons or positrons), 
respectively, and $\theta(i,j)$ the angles between the particles $i$ and
$j$ in the c.m.s. Furthermore, $m(q,q')$ denotes
the invariant mass of a quark pair $qq'$. 

We perform a number of checks.
The matrix elements have been checked against {\tt
MADGRAPH} \cite{MADGRAPH} and the phase space integrals have been checked 
against their asymptotic limits obtained analytically.
The electromagnetic gauge invariance of the matrix element of 
the bremsstrahlung process has been checked numerically and for some 
final states also analytically.

The cut independence of the total bremsstrahlung cross section
$\sigma_{\gamma}=\sigma_s + \sigma_h$ has been tested. $\sigma_s$
denotes the soft photon contribution to the cross section, which
includes photons of energy $E_{\gamma} \leq \omega$, and $\sigma_h$ is
the corresponding hard bremsstrahlung cross section for photons of
energy $E_{\gamma} > \omega$. Typical\footnote{i.e., representatives of the
classes I, II and III of processes considered below (see Tab.~ 2)}
results are presented in Table~1, where we have used parameters the
Eqs.~(\ref{masses1}), (\ref{coups2}) and (\ref{mlepton}).
As the infrared (IR) singular virtual one-loop corrections have not been
included, $\sigma_{{\rm s}}$ depends on a small fictitious
photon mass $m_\gamma$ which has been chosen to be 
$m_\gamma = 10^{-6}\, \gv$.  
The results in Table~1 are in a sense a measure of the numerical stability 
of our calculation.

\begin{center}
Table~1: Bremsstrahlung cross sections $\sigma_s$, $\sigma_h$ and
$\sigma_s+\sigma_h$ in fb for two different photon energy cuts $\omega$
and two c.m.s. energies of selected four--fermion reactions. No other 
kinematical cuts are imposed. 
The photon mass is $m_{\gamma}=10^{-6}GeV$.\\[2mm]
\begin{tabular}{|l|c|c|c|c|c|c|c|}
\hline
\rule{0mm}{7mm} Final state  & $\omega$   
      & \multicolumn{3}{|c }{$\sqrt{s}=190$ GeV} 
                    & \multicolumn{3}{|c|}{$\sqrt{s}=500$ GeV} \\ [2mm]
\cline{3-8}
 &  (GeV)   & \rule{0mm}{6mm}$\sigma_s$ 
& $\sigma_h$ & $\sigma_s +\sigma_h$ 
               & $\sigma_s$ & $\sigma_h$ & $\sigma_s +\sigma_h$ \\[1.5mm]
\hline
\hline
\rule{0mm}{7mm}$\nu_{\tau}\tau^+e^-\bar{\nu}_e\gamma$ 
   & $10^{-3}$ & 60.60(4) & 420.9(8) & 481.5 
                          & 24.01(4) & 683.6 (2.2) &  707.6 \\
   & $10^{-1}$ & 258.1(2) & 223.8(4) & 481.9 
                          & 305.3(3) & 403.4(1.2)  &  708.7 \\[2mm]
$\nu_{\tau}\tau^+\mu^-\bar{\nu}_{\mu}\gamma$ 
 & $10^{-3}$ & 63.26(3) & 314.6(4) & 377.9 & 13.74(1) & 155.8(3) & 169.5\\
 & $10^{-1}$ & 210.8(1) & 167.2(2) & 378.0 & 73.46(4) &  96.0(2) & 169.5\\[2mm]
$\nu_{\mu}\bar{\nu}_{\mu}\tau^-\tau^+\gamma$ 
   & $10^{-3}$ & 2.806(1) & 13.97(2) & 16.78 
                          & 0.8071(6) & 6.193(11) & 7.001 \\
   & $10^{-1}$ & 9.171(4) & 7.609(11) & 16.78 
                          & 3.158(2)  & 3.841(7)  &  6.999 \\[1.5mm]
\hline
\end{tabular}
\end{center}

Our results for the fermion mass effects in $e^+ e^- \ra 4f$ and $e^+
e^- \ra 4f\gamma$ in the presence of the cuts specified in
Eq. (\ref{cancuts}) are shown for several four--fermion final states
in Table 2, where we have used the parameters of
Eqs.~(\ref{masses2}--\ref{mquark}).  In columns 3 and 4, we list
respectively the results of Refs.~\cite{EXCALIBUR} and ~\cite{DDRW1}
which were obtained in the massless fermion limit. The results of our
calculation with non--zero fermion masses are shown in column
5. Wherever there is a substantial difference between the results of
Refs.~\cite{EXCALIBUR} and ~\cite{DDRW1} and the present work, we
present, in column 5, an additional entry representing the result of
ours obtained with the initial fermion masses equal to zero and the
final quark and/or charged lepton masses equal to $m_e$. We do not use
an exact zero mass limit for the final state fermions as it is not
easy to implement it in our kinematics routine, especially for the
bremsstrahlung reactions. However, with the cuts of
Eq.~(\ref{cancuts}) the tiny value of $m_e$ should not play a
numerically relevant role for the final state particles.  With this
substitution for the fermion masses, our results agree nicely with
those of Refs.~\cite{EXCALIBUR} and ~\cite{DDRW1}, except for $\epm
\ra \nu_{\tau}\bar{\nu}_{\tau}\mu^-\mu^+\gamma$, where the difference
amounts to a few standard deviations. We do not know the reason for
this discrepancy.  It may come from the fact that it is not possible
to approach the exact zero mass limit in the inverse of the invariant
mass of the $\mu^+\mu^-$ pair and the $\sim 1/s_{\mu+\mu-}$ behaviour
of the squared matrix element of $\epm \ra
\nu_{\tau}\bar{\nu}_{\tau}\mu^-\mu^+(\gamma)$ has to be mapped away in
order to improve a convergence of the phase space integral.  Actually
{\tt EXCALIBUR} crashes if there is no lower cut on $s_{\mu+\mu-}$.
The lower the cut, the greater the cross section. In our case there is
a natural lower cut on $s_{\mu+\mu-}$ of $4 m_{\mu}^2$ or $4 m_e^2$ in
case of the second entry in column 5. Therefore, we suppose that there
has been a tiny cut on the invariant mass of the $\mu^+\mu^-$ pair
imposed in Ref.~\cite{DDRW1} despite the fact that it was not
specified in the paper. For the non--radiative channels also a
comparison with {\tt KORALW}~\cite{KORALW} has been performed. Results
agree within the Monte Carlo errors.

The final states presented in Table~2 can be classified into three different
classes I -- III. According to classification of Ref.~\cite{BBLOR} class
I corresponds to the CC20 family, class II to the CC11 family and class 
III includes the leptonic processes of the NC32 family.
We see almost no mass effect for the tree level four--fermion reactions
of class I and II. However, there is a difference of $\sim 1$\% between
cross sections of the radiative reactions 
$\epm \ra u\bar{d}e^-\bar{\nu}_e\gamma$ and
$\epm \ra c\bar{s}e^-\bar{\nu}_e\gamma$ as well as between the results
for $\epm \ra u\bar{d}s\bar{c}\gamma$ obtained in the mass limit and
with the non zero fermion masses. The effect is even larger $(\sim 2 \%)$
for $\epm \ra \nu_{\tau}\tau^+e^-\bar{\nu}_e\gamma$ and 
$\epm \ra \nu_{\tau}\bar{\nu}_{\tau}\mu^-\mu^+\gamma$.

\begin{center}
Table~2: Mass dependence of cross sections at $\sqrt{s}=190$ GeV (in fb) 
for three different classes of final states with the cuts of 
Eq.~(\ref{cancuts}).
The second entries in column 5 correspond to the initial fermion
masses equal to zero and the final quark and/or charged lepton masses
equal to $m_e$.

\medskip

\begin{tabular}{|c|l|c|c|cc|}
\hline
\rule{0mm}{7mm} $\sigma$ & Final state & Ref.~\cite{EXCALIBUR} &Ref.~\cite{DDRW1} 
                           & \multicolumn{2}{c|}{Present work} \\
                (fb)   & of $\epm \ra$              & 
\multicolumn{2}{c|}{$(m_{\rm f}=0)$}   
                           & \multicolumn{2}{c|}{$(m_{\rm f} \neq 0)$} \\[2mm]
\hline
\rule{0mm}{6mm} 1 & \hspace*{8mm} 2 & 3 & 4  & \multicolumn{2}{c|}{5} \\[1mm]
\hline
\rule{0mm}{7mm}& $u\bar{d}e^-\bar{\nu}_e$ &691.5(8)& 693.6(3) & 693.4(6) &   \\
               & $c\bar{s}e^-\bar{\nu}_e$ & -- & -- & 693.1(6) &   \\[2mm]
 &$u\bar{d}e^-\bar{\nu}_e\gamma  $   & -- &   220.8(4)    & 220.3(7) &   \\
I &$c\bar{s}e^-\bar{\nu}_e\gamma $   & -- &     --        & 218.2(7) &   \\[2mm]
 &$\nu_{\mu}\mu^+e^-\bar{\nu}_e  $   & 227.0(3) &   227.5(1)    & 227.5(2) &   \\
 &$\nu_{\tau}\tau^+e^-\bar{\nu}_e$   & -- &     --        & 227.3(2) &   \\[2mm]
 &$\nu_{\mu}\mu^+e^-\bar{\nu}_e\gamma$& -- &    79.1(1)    &  79.0(3) &   \\
 &$\nu_{\tau}\tau^+e^-\bar{\nu}_e\gamma$& -- &     --      &  77.5(2) & \\[1.5mm]
\hline
\rule{0mm}{7mm} & $u\bar{d}\mu^-\bar{\nu}_{\mu}$       
          & 667.4(8)      &   666.7(3)    &     666.7(4) &   \\
 &$u\bar{d}\tau^-\bar{\nu}_{\tau}$ & --  &       --      &     666.0(3) &   \\[2mm]
 &$u\bar{d}\mu^-\bar{\nu}_{\mu}\gamma$ & -- &   214.5(4)  &     213.8(3) &   \\
II  &$u\bar{d}\tau^-\bar{\nu}_{\tau}\gamma$ & -- &   --   &     209.3(5) &   \\[2mm]
 &$\nu_{\tau}\tau^+\mu^-\bar{\nu}_{\mu}$ 
                      & 218.7(3)     &   218.6(1)   & 218.3(1) & 218.5(1)\\
 &$\nu_{\tau}\tau^+\mu^-\bar{\nu}_{\mu}\gamma$ 
                                  & --  &    76.7(1)    & 75.1(2)& 76.6(2)\\[2mm]
 &$u\bar{d}s\bar{c}$         & -- &  2015.3(8)    &    2016(1)&      \\
 &$u\bar{d}s\bar{c}\gamma$
      & --  &    598(1)     &  593(2)& 598(1) \\[1.5mm]
\hline
\rule{0mm}{7mm} & $\tau^-\tau^+\mu^-\mu^+$ 
                        & --  &    11.02(1)   &   9.26(1) & 11.03(2)\\
 &$\tau^-\tau^+\mu^-\mu^+\gamma$ & --  &     6.78(3)   &   5.32(3) 
                                                     &  6.62(5)\\[2mm]
 &$\nu_{\tau}\bar{\nu}_{\tau}\mu^-\mu^+$ & 10.121(40)&  10.103(8)  &   10.05(1) 
                                                     & 10.095(10)\\
III &$\nu_{\mu}\bar{\nu}_{\mu}\tau^-\tau^+$ & -- &  --  &  8.529(6) &   \\[2mm]
 &$\nu_{\tau}\bar{\nu}_{\tau}\mu^-\mu^+\gamma$ & -- & 4.259(9) & 4.18(2)  & \\
 &$\nu_{\mu}\bar{\nu}_{\mu}\tau^-\tau^+\gamma$ & -- &  --   &  3.167(7)   & \\[2mm]
 &$\nu_{\tau}\bar{\nu}_{\tau}\nu_{\mu}\bar{\nu}_{\mu}$& 8.224(6) &  8.218(2)  & 8.222(5) 
                                                     &  \\
 &$\nu_{\tau}\bar{\nu}_{\tau}\nu_{\mu}\bar{\nu}_{\mu}\gamma$ 
                & -- & 1.511(1) & 1.510(3) & \\[1.5mm]
\hline
\end{tabular}
\end{center}

The fermion mass effects become more pronounced for the reactions belonging
to class III. We observe it here already for the tree level reactions
$\epm \ra\tau^-\tau^+\mu^-\mu^+$ and 
$\epm \ra \nu_{\mu}\bar{\nu}_{\mu}\tau^-\tau^+$ where it amounts to
about 15\% and it becomes even stronger for the corresponding bremsstrahlung
reactions. The reason for that is the presence of the virtual photon
exchange graphs which introduce the $\sim 1/s$ behavior of the matrix
element and the lack of the $W^{\pm}$-boson exchange
graphs which are relatively insensitive to the fermion masses. 

\section{Summary}

We have studied the fermion mass effects for several channels of 
$e^+ e^- \ra 4f$ and $e^+ e^- \ra 4f\gamma$ using the method of calculation 
elaborated in Refs.~\cite{KZ} and \cite{JK}.
It has been shown that the fermion mass effects may affect the desired 
accuracy 
of 0.5\% for the final LEP2 data analysis and the expected precision of
0.1\% for the linear collider even in the presence of cuts. Therefore, it
seems to be much better not to neglect the fermion masses in  
calculations intending to match the high precision of the present
and future experiments.

The cut independent results presented in Table~1 may serve
as tests of MC generators which work with massless fermions especially
in the collinear regions of phase space.

\bigskip
{\bf Acknowledgment}

One of us (K.K.) would like to thank DESY Zeuthen for warm hospitality
during his stay in September, 2000. We thank S. Jadach for providing us
results of KORALW and Ch. Ford for carefully reading the manuscript.

\bigskip

\end{document}